\documentclass{aa}

\usepackage{graphicx}
\graphicspath{{Figs/}}
\usepackage{txfonts}
\usepackage[colorlinks=true,citecolor=blue]{hyperref}

\begin{document}

   \title{Nine new eclipsing accreting ultracompact white dwarf binaries from the Zwicky Transient Facility}

   \author{J. van Roestel\inst{1,2}\corrauth{jcjvanroestel.astro@gmail.com}
        \and M.~J.~Green\inst{3,4}
        \and T.~L.~S.~Wong\inst{5}
        \and P.~Szkody\inst{6}
        \and J.~M.~Khalil\inst{7}
        \and A.~Householder\inst{8,9}
        \and K.~B.~Burdge\inst{10,8}
        \and T.~A.~Prince\inst{11}
        \and L.~Blomberg\inst{11}
        \and I.~Caiazzo\inst{1}
        \and K.~K.~Das\inst{11}
        \and N.~Yamaguchi\inst{11}
        \and A.~C.~Rodriguez\inst{12,11}
        \and Z.~P.~Vanderbosch\inst{13}
        \and \\ E.~C.~Bellm\inst{14}
        \and T.~X.~Chen\inst{15}
        \and R.~Dekany\inst{16}
        \and A.~J.~Drake\inst{11}
        \and M.~J.~Graham\inst{11}
        \and G.~Helou\inst{15}
        \and M.~M.~Kasliwal\inst{11}
        \and A.~A.~Mahabal\inst{11}
        \and R.~L.~Riddle\inst{16}
        \and B.~Rusholme\inst{15}
        }

   \institute{Institute of Science and Technology Austria, Am Campus 1,
              3400 Klosterneuburg, Austria
         \and Anton Pannekoek Institute for Astronomy, University of Amsterdam,
              1090 GE Amsterdam, The Netherlands
         \and Homer L. Dodge Department of Physics and Astronomy, University of
              Oklahoma, 440 W. Brooks Street, Norman, OK 73019, USA
         \and JILA, University of Colorado and National Institute of Standards
              and Technology, 440 UCB, Boulder, CO 80309-0440, USA
         \and The Observatories of the Carnegie Institution for Science,
              Pasadena, CA 91101, USA
         \and Department of Astronomy, University of Washington, Box 351580,
              Seattle, WA 98195, USA
         \and Dipartimento di Fisica e Astronomia ``Augusto Righi'',
              Universit\`a di Bologna, Via Gobetti 93/2, 40129 Bologna, Italy
         \and Kavli Institute for Astrophysics and Space Research,
              Massachusetts Institute of Technology, 77 Massachusetts Avenue,
              Cambridge, MA 02139, USA
         \and Department of Earth, Atmospheric and Planetary Sciences,
              Massachusetts Institute of Technology, Cambridge, MA 02139, USA
         \and Department of Physics, Massachusetts Institute of Technology,
              77 Massachusetts Avenue, Cambridge, MA 02139, USA
         \and Division of Physics, Mathematics and Astronomy, California
              Institute of Technology, 1200 E. California Blvd,
              Pasadena, CA 91125, USA
         \and Center for Astrophysics $|$ Harvard \& Smithsonian, 60 Garden
              Street, Cambridge, MA 02138, USA
         \and Hobby-Eberly Telescope, McDonald Observatory, 32 Mount Locke Road,
              Fort Davis, TX 79734, USA
         \and DIRAC Institute, Department of Astronomy, University of
              Washington, 3910 15th Avenue NE, Seattle, WA 98195, USA
         \and IPAC, California Institute of Technology, 1200 E. California
              Blvd, Pasadena, CA 91125, USA
         \and Caltech Optical Observatories, California Institute of
              Technology, Pasadena, CA 91125, USA
              }

   \date{Received ...; accepted ...}

  \abstract
   {  Accreting ultracompact white dwarf binaries (accreting UCBs) transfer
   helium-rich material at orbital periods below about 65\,min, and mainly divide
   into hydrogen-free AM~CVn binaries and helium cataclysmic variables.
   Eclipsing systems are of particular value: they allow the component masses
   and radii to be measured essentially independently of stellar models, the most direct test of the
   proposed formation channels,
   and they form a sample with a simple, well-understood selection function.
   Only 21 eclipsing accreting white dwarf UCBs have been published.}
   {We aim to enlarge the sample of eclipsing accreting UCBs available for
   detailed characterisation.}
   {We searched the Zwicky Transient Facility (ZTF) light curves of \textit{Gaia}-selected candidates and of ZTF-selected dwarf novae for short-period eclipse signals using a box least-squares period search. Keck/LRIS identification spectra were obtained for nine systems.}
   {We present nine new discoveries and re-analyse the four systems first
   reported by Khalil et al. (2024). Eclipses are detected directly in the
   survey photometry of all 13 systems, which have orbital periods between
   26.0 and 65.0\,min and median ZTF magnitudes $g = 18.7$--$21.5$. Ten are
   consistent with being classical AM~CVn binaries, and two are likely helium
   cataclysmic variables, of which one may host a high-state disc. The
   remaining system, ZTF~J2116+2446, shows only hydrogen lines, making it the
   shortest-period UCB with hydrogen but no detected helium; it may have
   evolved from a white dwarf--brown dwarf binary.}
   {The nine new systems increase the number of published eclipsing accreting
   UCBs from 21 to 30. Because they are eclipsing, they are immediately
   available for the high-speed photometric modelling that yields
   model-independent masses and radii, to be presented in future work,
   including for the three hydrogen-rich systems that are key tests of the
   evolved cataclysmic-variable channel.}


   \titlerunning{Eclipsing accreting UCBs from ZTF}
   \authorrunning{van Roestel et al.}

   \maketitle

\nolinenumbers

\section{Introduction}
\label{sec:intro}
Accreting ultracompact white dwarf binaries (accreting UCBs) are mass-transferring binaries in which a white dwarf (WD) accretes material from a compact donor (usually degenerate or semi-degenerate, but in a few systems a hot subdwarf or a warm, non-degenerate star), at orbital periods shorter than about 65\,min \citep[for reviews, see][]{2005ASPC..330...27N,2010PASP..122.1133S,2018A&A...620A.141R,green2025}. They are the tightest known interacting binaries with a white dwarf accretor and are of astrophysical interest for various reasons: they are laboratories for accretion physics and for the binary evolution of stellar remnants, and are among the strongest predicted Galactic gravitational-wave sources in the millihertz band. Some of the known systems will be detectable by the Laser Interferometer Space Antenna \citep[LISA;][]{kupfer2018,kupfer2024,amaroseoane2023}. 

One of the main open questions about UCBs is how these binary systems form. Three channels have been proposed. In the double-white-dwarf channel, a detached pair of white dwarfs formed after common-envelope phases is driven into contact by gravitational-wave radiation. Mass transfer begins from a (semi-)degenerate, low-mass helium-core white dwarf donor at the shortest orbital periods of the class, below the $\sim$10\,min limit where the accretion stream impacts the accretor directly rather than forming a disc. The orbit shrinks to a period minimum of only a few minutes before widening again as the donor is drained of mass \citep[e.g.][]{1979AcA....29..665T,1981ApJ...244..269N,2007MNRAS.381..525D,2006ApJ...640..466B,2021ApJ...923..125W,2022ApJ...935....9C,chakraborty2024}. In the helium-star channel, the donor is instead a non-degenerate, helium-burning star that is brought into contact by the same gravitational-wave-driven angular-momentum loss, beginning mass transfer at a period of several tens of minutes and shrinking to a period minimum of about 10\,min, at which point helium burning ceases and the donor becomes semi-degenerate. As in the double-white-dwarf channel, the orbit then widens again as the donor continues to lose mass \citep[e.g.][]{1986A&A...155...51S,1991ApJS...76...55I,1987ApJ...317..717I,2008AstL...34..620Y,2021ApJ...922..245B,2023MNRAS.519.2567S,2025A&A...704A..82R}. In the evolved cataclysmic-variable (CV) channel, the donor begins as an ordinary, hydrogen-rich, near-main-sequence star already transferring mass at a much longer orbital period, and only becomes helium-enriched once nuclear evolution and continued mass loss have stripped away most of its hydrogen envelope \citep[e.g.][]{1985PAZh...11..123T,2003MNRAS.340.1214P,2023MNRAS.520.3187S}. For a long time, the double-white-dwarf and helium-star channels were the most favoured from population-synthesis and composition arguments \citep[e.g.][]{nelemans2004,nelemans2010,2015ApJ...809...80G}, but recently, \citet{2023A&A...678A..34B} have argued that the evolved CV channel could explain at least a significant fraction of the population, and \citet{2025A&A...697A.100B} have used it to explain a specific system that was previously difficult to reconcile with the other two channels.

More than a hundred accreting UCBs are now known \citep{green2025}. This sample shows a rich phenomenological diversity. \citet{green2025} divides them into two observational categories: AM~CVn binaries, which show no hydrogen in their optical spectra, and helium cataclysmic variables (He~CVs), which show hydrogen but are helium-enriched relative to the general CV population. They can also be grouped by their accretion behaviour, which correlates strongly with orbital period. Very-short-period systems ($\lesssim 10$\,min) are direct-impact accretors and tend to be bright X-ray sources \citep{marsh2002}. Systems with periods of 10--22\,min have an accretion disc in a constant high state \citep{tsugawa1997} and are typically very blue and bright in the UV \citep{2018A&A...620A.141R}. Longer-period systems show optical outbursts (dwarf novae) that become less frequent as the period increases, with recurrence times exceeding 10\,yr at orbital periods of $\approx 40$\,min \citep{levitan2015}. However, a growing number of UCBs deviate from this picture \citep{rivera-sandoval2021,2021RNAAS...5....3W,kato2023,chakraborty2024,chickles2025,kosakowski2025}. This also includes the (spectroscopic) class of He~CVs, which tend to have long periods but do show frequent outbursts \citep{2020MNRAS.496.1243G}.

A few accreting white dwarf binaries stand apart because of the nature of their donor. The eclipsing WD+subdwarf-B binaries \object{ZTF~J2130+4420} \citep{2020ApJ...891...45K} and \object{ZTF~J2055+4651} \citep{2020ApJ...898L..25K} accrete from a Roche-lobe-filling hot subdwarf, and \object{ZTF~J1813+4251} \citep{2022Natur.610..467B} and \object{ZTF~J0127+5258} \citep{burdge2023} from a warm donor. These systems suggest that we have not yet uncovered the full observational diversity of the UCB population.

Because of their phenomenological diversity, accreting UCBs have been found by a range of methods: detection of dwarf-nova outbursts \citep[e.g.][]{levitan2015,breedt2014,2026PASA...43...52K}, colour-based selection \citep[e.g.][]{1957ApJ...126...14G,roelofs2009,2013MNRAS.429.2143C}, X-ray luminosity or variability \citep[e.g.][]{israel1999,rodriguez2023}, spectroscopic detection of helium emission \citep[e.g.][]{anderson2005,2026arXiv260722836I}, and detection of eclipses in low-cadence survey photometry or high-speed photometry \citep[e.g.][]{vanroestel2022,chakraborty2024}. This diversity and the large differences in detection methods and biases make it challenging to study the intrinsic properties of the UCB population. 

For systematic population studies, eclipsing systems have several advantages. First of all, with the present-day abundance of deep photometric data, eclipsing UCBs can be identified with just photometry, which also directly yields the orbital period. In addition, they are easier to characterise. Modelling the eclipse profile of a high-speed light curve \citep[e.g.][]{1986MNRAS.219..629W,2010MNRAS.402.1824C,green2018,vanroestel2022} yields the mass ratio, the inclination, and the stellar radii in units of the orbital separation, and hence the component masses and radii directly with only an assumed mass--radius relationship for the accretor. Eclipses also allow changes in the orbital period to be measured, which probes the effect of gravitational waves and accretion physics \citep{2023MNRAS.518.5123M,chakraborty2024,2026MNRAS.548ag673G}. Finally, an eclipse-selected sample has a selection function simple enough to model (set by the inclination-dependent eclipse probability and the survey's depth and cadence), whereas samples selected on spectroscopic follow-up or on outburst activity carry biases that are far harder to quantify. The properties of the sample of eclipsing UCBs can therefore be corrected for selection effects to recover intrinsic population properties, such as the space density and period distribution, that can be compared with the predictions of population-synthesis models \citep[e.g.][]{2001A&A...368..939N,2015ApJ...809...80G}.

The first two discovered accreting UCBs undergoing eclipses were \object{YZ~LMi} \citep{anderson2005,copperwheat2011} and \object{Gaia14aae} \citep{campbell2015,green2018}. Systematic searches of the Zwicky Transient Facility \citep[ZTF;][]{bellm2019,graham2019,masci2019} and of ATLAS \citep{tonry2018} have since greatly increased the known sample \citep{burdge2020,vanroestel2022,rodriguez2023,khalil2024,chakraborty2024,chickles2026}. Counting the two WD+subdwarf-B binaries and the two warm-donor systems ZTF~J1813+4251 and ZTF~J0127+5258 alongside the AM~CVn binaries and He~CVs, there are currently 21 eclipsing accreting white dwarf UCBs published\footnote{An up-to-date overview of all eclipsing accreting white dwarf UCBs is maintained at \url{https://doi.org/10.5281/zenodo.22706718}.}.

In this paper we report the discovery of nine further eclipsing accreting UCBs (two of which are published simultaneously by \citealt{2026arXiv261040346H}), and include the four eclipsing systems first reported by \citet{khalil2024}, which were selected in the same way; we analyse all 13 together. 

\section{Sample selection and observations}

\subsection{Target search}
\label{sec:search}

The 13 systems were identified with the same methods used in our previous searches \citep{vanroestel2022,khalil2024}, but with a broader initial data set. We assembled candidate ultracompact binaries from two sets of ZTF light curves: all \textit{Gaia}-selected white dwarf candidates \citep[the \textit{Gaia} EDR3 white dwarf catalogue of][]{gentilefusillo2021}, and all ZTF-selected dwarf novae \citep{2020AJ....159..198S,khalil2024}. The dwarf-nova candidates were selected using ZTF alert photometry from the Kowalski data archive \citep{2019JOSS....4.1247V,2023ApJS..267...31C}. The two selections recover different parts of the population. A system that never outbursts (typically with $P_\mathrm{orb} \gtrsim 40$\,min) sits close to the white dwarf sequence in colour and absolute magnitude, while a frequently outbursting system ($P_\mathrm{orb} \lesssim 40$\,min, including systems too faint to be detected by \textit{Gaia}) may sit above the white dwarf sequence, but is recovered by the dwarf-nova data set.
For each target, we used the ZTF forced photometry \citep{2023arXiv230516279M}. The forced photometry provides a flux measurement for each exposure regardless of signal-to-noise ratio\footnote{as opposed to alerts and PSF photometry, which do not include low SNR data}. This forced photometry allowed us to greatly expand our initial data set, and is the reason (together with the longer time baseline) for the substantial increase in the number of detected UCBs over our previous works.

We refer to \citet{vanroestel2022} for the details of the period search method. In brief, for each candidate we searched for periodic eclipses with the GPU implementation of the box least-squares algorithm \citep{kovacs2002} in \texttt{cuvarbase}, and inspected the phase-folded light curves of the strongest peaks by eye. Orbital periods and mid-eclipse epochs were then refined by fitting the folded photometry with a simple trapezoid-and-sine model, which reproduces the eclipse shape and any orbital-phase modulation without requiring a full physical light-curve model. The recovered ephemerides are listed in Table~\ref{tab:ephemerides}. All times are barycentric Julian dates on the barycentric dynamical time scale. During preparation of this paper, we became aware that two of our targets (\object{ZTF~J1302$-$1746} and \object{ZTF~J0859$-$1110}) were also found by another group \citep{2026arXiv261040346H} using a similar method, and are presented simultaneously in that work and in this one. In addition, \object{ZTF~J0151+0909} and ZTF~J0859$-$1110 were recently identified as AM~CVn binaries from their optical spectra, by \citet{2026arXiv260722836I} and \citet{2026arXiv260727855H}, respectively; neither work reported eclipses or an orbital period.

\subsection{Spectroscopy}

Identification spectra were obtained for nine of the 13 systems with the Low Resolution Imaging Spectrometer \citep[LRIS;][]{oke1995} on the Keck\,I telescope, and reduced with \texttt{lpipe} \citep{perley2019}. We used the D560 dichroic with the 600/4000 grism on the blue arm and either the 400/8500 or the 600/7500 grating on the red arm, through a 1\farcs0 slit, giving a resolution of $R \approx 1000$--$1500$ (FWHM $\approx 4$--$7$\,\AA) over the usable range of 3600--8200\,\AA. Individual exposure times ranged from 300 to 900\,s. Pixels flagged by the reduction were rejected before analysis. None of the spectra was taken during an outburst. The nine systems were observed on eleven nights in total, each of which is marked in Fig.~\ref{fig:outburst_lcs}: no night falls inside a detected outburst, and for ten of the eleven the survey photometry within 3\,d of the spectrum is at the quiescent level of that system. The exception is the second night of \object{ZTF~J0654+3608}, where the nearest photometry is 5\,d away, but this system does not outburst in general. Two spectra were taken close in time to an outburst but still at the quiescent level: \object{ZTF~J0701+5023} was observed 4.8\,d before the onset of an outburst, and \object{ZTF~J1747+6302} between two outbursts, 12\,d after the end of one and 11\,d before the start of the next. The remaining four systems are faint ($g > 20$) and have not yet been observed spectroscopically.

\section{Results}

\begin{table*}
\centering
\caption{Eclipsing accreting ultracompact binaries presented in this work. Orbital periods are the refined values from the trapezoid-and-sine fit to the survey photometry (Table~\ref{tab:ephemerides}). Reported magnitudes are the median of the ZTF forced-photometry light curve. Distances are the geometric estimates from \citet{bailerjones2021}, with the quoted uncertainties spanning their 16th to 84th percentile interval. $N_{\rm ob}$ is the number of outbursts identified in the ZTF light curve and $t_{\rm rec}$ the raw recurrence time (the baseline divided by $N_{\rm ob}$, Sect.~\ref{sec:outbursts}). For $N_{\rm ob} = 1$ no interval between successive outbursts has been observed and $t_{\rm rec}$ is a lower limit; systems with no detected outburst are marked with a dash. The subclasses of these objects are discussed in Sect.~\ref{sec:nature}; a question mark denotes a tentative classification. The Lines column records whether hydrogen and helium are identified in the LRIS spectrum, and a dash means no spectrum is available (Sect.~\ref{sec:spectra}). The individual line transitions are listed in Table~\ref{tab:lines}.}
\label{tab:targets}
\renewcommand{\arraystretch}{1.2}
\begin{tabular}{lccccccr p{1.6cm} l}
\hline\hline
Name & R.A. & Dec. & $P_{\rm orb}$ & $g$ & $d$ & $N_{\rm ob}$ & $t_{\rm rec}$ & Lines & Subclass \\
     & (J2000) & (J2000) & (min) & (AB mag) & (pc) & & (yr) & &  \\
\hline
ZTF~J0729$-$0602\tablefootmark{a} & 07:29:07.68 & $-$06:02:46.6 & 26.0 & 20.50 & $2029^{+3613}_{-930}$ & 13 & 0.6 & He & AM~CVn \\
ZTF~J0614+6014 & 06:14:28.27 & +60:14:38.0 & 31.1 & 21.51 & -- & 6 & 1.3 & -- & AM~CVn \\
ZTF~J0701+5023\tablefootmark{a} & 07:01:15.77 & +50:23:21.5 & 32.5 & 20.38 & $1637^{+1290}_{-669}$ & 8 & 1.0 & He & AM~CVn \\
ZTF~J0555$-$2345\tablefootmark{b} & 05:55:02.52 & $-$23:45:06.5 & 34.5 & 20.63 & $2262^{+1913}_{-1113}$ & 2 & 3.9 & -- & AM~CVn \\
ZTF~J1302$-$1746\tablefootmark{c,d} & 13:02:07.68 & $-$17:46:41.9 & 39.3 & 19.63 & $574^{+157}_{-111}$ & 1 & $>$8.3 & He & AM~CVn \\
ZTF~J0654+3608 & 06:54:29.52 & +36:08:53.9 & 39.3 & 20.91 & -- & 0 & -- & H, He? & He~CV? \\
ZTF~J1838+0744\tablefootmark{a} & 18:38:47.15 & +07:44:46.3 & 39.6 & 20.33 & $1741^{+1693}_{-851}$ & 4 & 2.1 & He & AM~CVn \\
ZTF~J0151+0909\tablefootmark{e} & 01:51:15.50 & +09:09:49.0 & 44.7 & 18.69 & $233^{+12}_{-10}$ & 0 & -- & He & AM~CVn \\
ZTF~J2116+2446\tablefootmark{a} & 21:16:04.75 & +24:46:20.6 & 56.1 & 21.41 & -- & 3 & 2.8 & H & CV? \\
ZTF~J0859$-$1110\tablefootmark{c,f} & 08:59:34.30 & $-$11:10:55.5 & 59.9 & 19.49 & $199^{+15}_{-10}$ & 0 & -- & He & AM~CVn \\
ZTF~J1146$-$1649 & 11:46:21.06 & $-$16:49:18.4 & 61.0 & 20.50 & $604^{+405}_{-199}$ & 0 & -- & -- & AM~CVn \\
ZTF~J0531$-$2019\tablefootmark{d} & 05:31:21.10 & $-$20:19:39.7 & 64.6 & 20.88 & -- & 1 & $>$7.9 & -- & AM~CVn? \\
ZTF~J1747+6302 & 17:47:25.57 & +63:02:47.8 & 65.0 & 21.00 & -- & 25 & 0.3 & H, He & He~CV \\
\hline
\end{tabular}
\renewcommand{\arraystretch}{1}
\tablefoot{
\tablefoottext{a}{Eclipses first reported by \citet{khalil2024}.}
\tablefoottext{b}{Previously reported as an optical transient, MASTER OT J055502.52$-$234506.5, by the MASTER network \citep{2016ATel.8846....1V}.}
\tablefoottext{c}{Independently discovered by \citet{2026arXiv261040346H}, published simultaneously with this work.}
\tablefoottext{d}{Previously known as a dwarf nova from the Catalina Real-time Transient Survey \citep[CRTS;][]{2009ApJ...696..870D}.}
\tablefoottext{e}{Recently identified as an AM~CVn binary from its optical spectrum by \citet{2026arXiv260722836I}.}
\tablefoottext{f}{Recently identified as an AM~CVn binary from its optical spectrum by \citet{2026arXiv260727855H}.}
}
\end{table*}

The 13 systems are listed in Table~\ref{tab:targets}, including the four systems reported by \citet{khalil2024}. Their orbital periods span 26.0 to 65.0\,min, and their median ZTF magnitudes range from $g = 18.7$ to $21.5$. Eight of the 13 have a geometric distance derived from the \textit{Gaia} DR3 parallax \citep{gaiadr3,bailerjones2021}; four of those lie between about 200 and 600\,pc, while the remaining four are nominally at 1.6--2.3\,kpc but very poorly constrained. The other five systems have no useful parallax.

\subsection{Eclipses in the ZTF photometry}

\begin{figure*}[t]
    \centering
    \includegraphics[width=\textwidth]{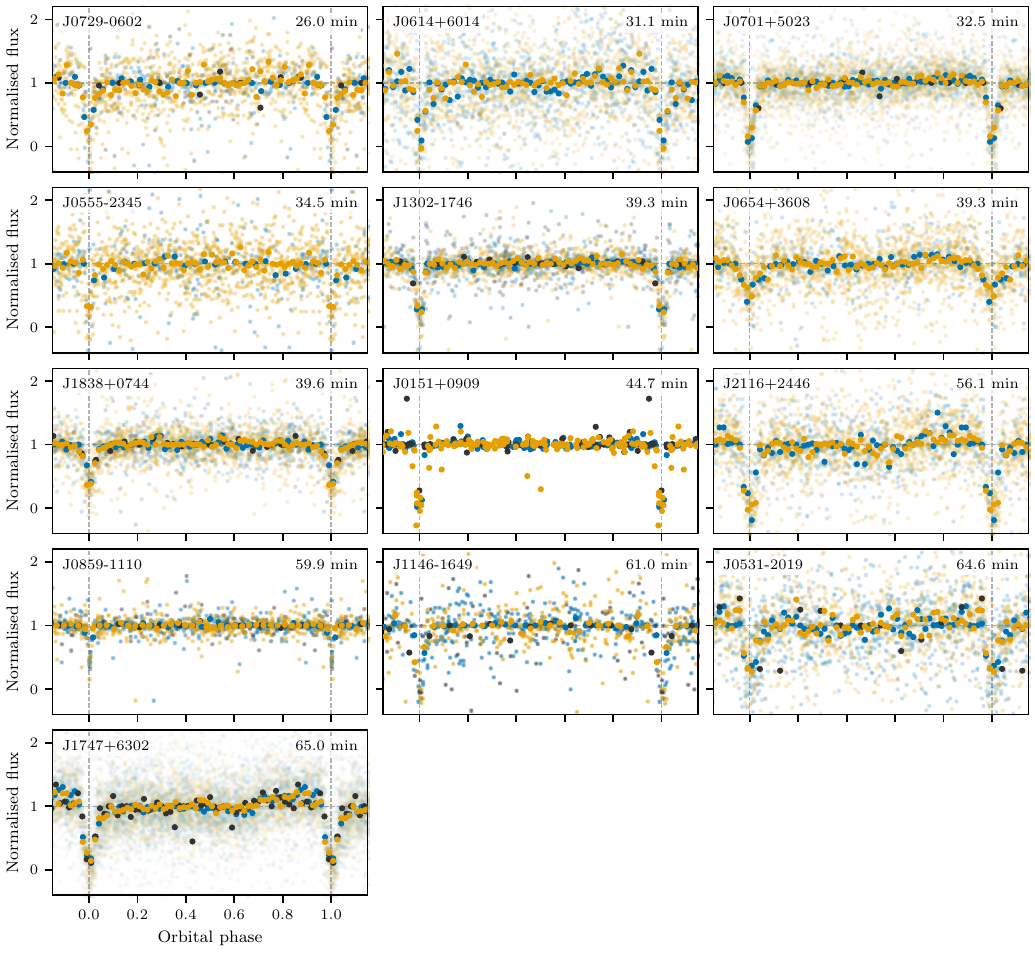}
    \caption{Discovery light curves of the 13 eclipsing accreting UCBs, folded on the orbital period. Faint points are the individual epochs; the larger dots are the median-binned light curves (ZTF $g$, $r$, $i$ in blue, orange, and dark grey). For light curves with $\le 500$ epochs, the binned data are omitted and individual epochs are plotted opaquely. All panels share the same flux scale, normalised to the out-of-eclipse level. Bands with fewer than 40 epochs are not shown. Outbursts reach fluxes well above the plotted range and are not shown.}
    \label{fig:discovery_lcs}
\end{figure*}

Figure~\ref{fig:discovery_lcs} shows the ZTF light curves of all 13 systems, folded on the orbital period, with the eclipse at phase 0. The depths of the binned folded curves range from about 20\% to essentially total (100\%). Most of these systems are close to the ZTF detection limit and therefore individual measurements are noisy, but because of the large number of epochs, the periodic signal of all these objects is clearly detected. In some systems, a clear pre-eclipse brightening is detected, caused by the accretion hot spot rotating into view, which is typical for accreting white dwarfs \citep[see e.g.][]{2019MNRAS.486.5535M}. 

\subsection{Spectra}
\label{sec:spectra}
\begin{figure*}[t]
    \centering
    \includegraphics[width=\textwidth]{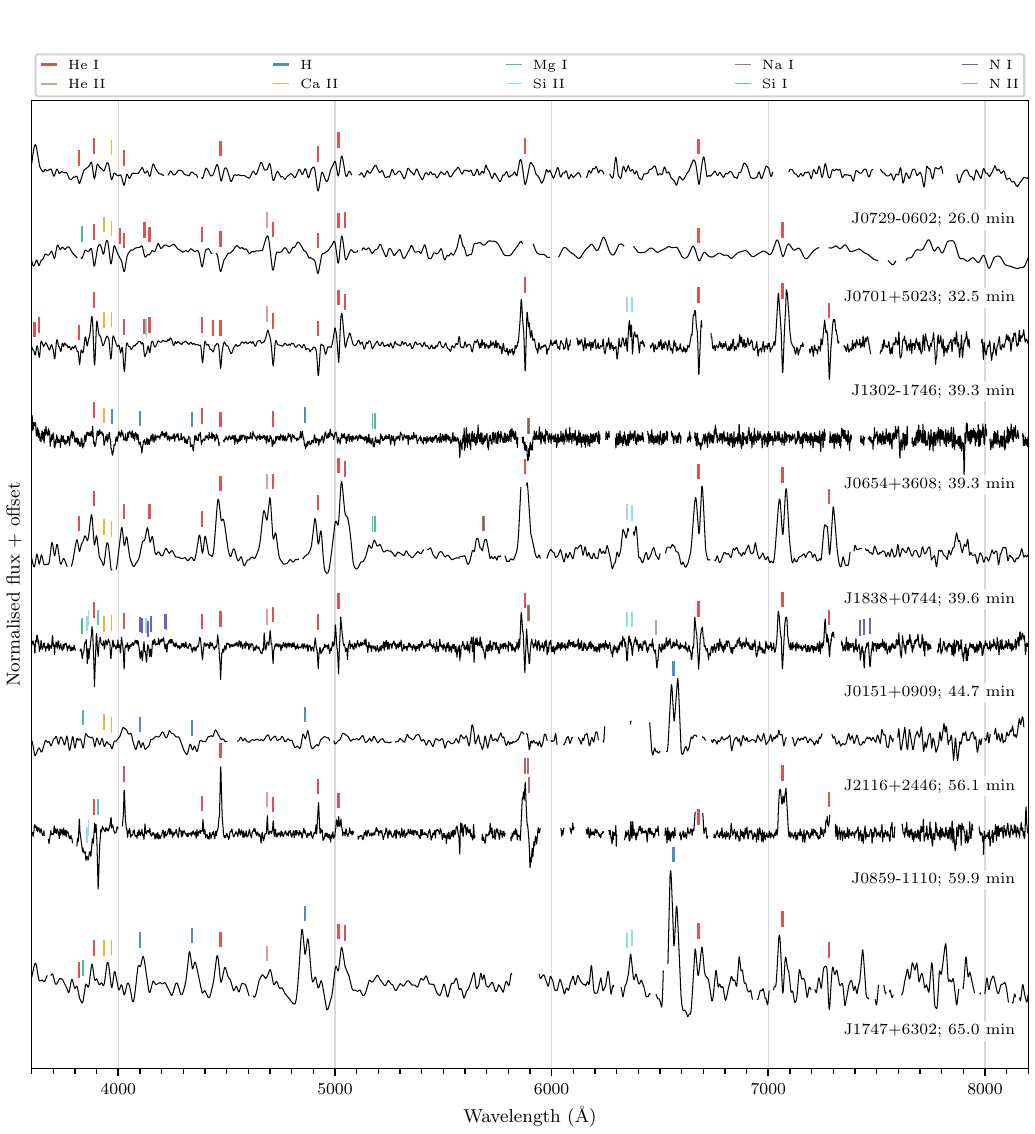}
    \caption{Keck/LRIS identification spectra of nine of the 13 systems,
    ordered by orbital period with the longest period at the bottom, and offset
    vertically for clarity. Each spectrum has been divided by a running median to remove the continuum slope and smoothed with a Gaussian kernel whose width is set by its S/N. Coloured ticks mark the identified lines, also listed in Table~\ref{tab:lines}.}
    \label{fig:spectra}
\end{figure*}

The nine available LRIS spectra are shown in Fig.~\ref{fig:spectra}. We searched each for the species typically detected in AM~CVn spectra \citep[e.g.][]{kupfer2016,vanroestel2022}. Every candidate line was confirmed or rejected by eye on the continuum-divided spectra; confirmed lines are marked in Fig.~\ref{fig:spectra} and tabulated in Table~\ref{tab:lines}.

Six of the systems show spectra typical of long-period AM~CVn binaries ($\gtrsim 22$\,min): \ion{He}{i} lines from the helium-dominated accretion disc and no sign of hydrogen. The \ion{He}{i} lines typically consist of double-peaked disc emission, sometimes with an absorption core from the white dwarf itself. In some cases, metal absorption lines from the white dwarf (\ion{Ca}{ii}, \ion{Na}{i}, \ion{Mg}{i}), or emission lines from the disc (e.g. \ion{Si}{ii}) are also visible. The objects are \object{ZTF~J0729$-$0602}, ZTF~J0701+5023, ZTF~J1302$-$1746, \object{ZTF~J1838+0744}, ZTF~J0151+0909, and ZTF~J0859$-$1110.

Three objects show more unusual spectra. First of all, ZTF~J0654+3608 shows a spectrum with only absorption lines: \ion{Na}{i} and the hydrogen Balmer series are the strongest, with weaker \ion{He}{i} lines. This is unusual, and suggests that the disc is hot and in a high state, which is not typically seen at a period of 39\,min.

\object{ZTF~J2116+2446} and ZTF~J1747+6302 are two objects that show a strong, double-peaked H$\alpha$ emission line, accompanied by other, weaker Balmer lines. This shows that the donors in these systems still retain some hydrogen. ZTF~J2116+2446 does not seem to show any sign of \ion{He}{i}, while ZTF~J1747+6302 does show both \ion{He}{i} and \ion{He}{ii} lines.

\subsection{Outbursts}
\label{sec:outbursts}
Outbursts were identified and verified by eye on the unfolded light curves, assisted by a basic peak-finding method. Nine of the 13 systems show outbursts. The unfolded survey light curves illustrating these outbursts are shown in Fig.~\ref{fig:outburst_lcs} in Appendix~\ref{app:outburst_lcs}.

We calculate the raw recurrence times $t_{\rm rec}$ (the baseline $T$ divided by the number of outbursts $N_{\rm ob}$) in Table~\ref{tab:targets}, which range from 0.3 to 3.9\,yr for the seven systems with more than one detected outburst. ZTF~J1302$-$1746 and \object{ZTF~J0531$-$2019} show a single outburst each, for which no interval between successive events has been observed and $t_{\rm rec}$ is only a lower limit. The remaining four systems, ZTF~J0654+3608, ZTF~J0151+0909, ZTF~J0859$-$1110, and \object{ZTF~J1146$-$1649}, show no outburst at all over baselines of 7.9 to 8.4\,yr.

\section{Discussion}
\label{sec:discussion}

\subsection{The nature of the objects}
\label{sec:nature}
Here, we briefly discuss the nature of the UCBs in this sample, and how they compare with the previously discovered UCBs.

\subsubsection{Two hydrogen-rich systems: ZTF~J2116+2446 and ZTF~J1747+6302}
Two objects, ZTF~J2116+2446 and ZTF~J1747+6302, stand out from the rest of the sample because of the clear presence of hydrogen in their spectra (Sect.~\ref{sec:spectra}). In both objects, both H$\alpha$ and H$\beta$ are clearly detected, and there is no doubt that the accretion disc, and therefore the donor, still contains some hydrogen. We note that no helium is detected in ZTF~J2116+2446, while ZTF~J1747+6302 does show \ion{He}{i} emission lines. Finally, the primary eclipse is quite deep in both systems, which suggests that the donor contributes little to no light in the optical (as opposed to ZTF~J1813+4251, which has a warm, optically detectable donor; see Sect.~\ref{sec:intro}).

Both objects also stand out because of their outburst behaviour (see Fig.~\ref{fig:outburst_lcs}). At periods of $P_{\rm orb} = 50$--$70$\,min AM~CVn discs are expected to be stable for decades between outbursts (\citealt{levitan2015}; but see Sect.~\ref{sec:intro}). At periods of 56.1 and 65.0\,min and recurrence times of 2.8 and 0.3\,yr, respectively, both objects clearly stand out from typical AM~CVn systems at these orbital periods, while being relatively typical for He~CVs.

These observational signatures suggest that these two objects have a different evolutionary history from the more common AM~CVn systems at these long orbital periods. The characteristics of ZTF~J1747+6302 are typical of He~CVs, and the evolved CV channel is the most commonly invoked explanation for He~CVs \citep[e.g.][]{2023A&A...678A..34B}.
ZTF~J2116+2446 is more unusual, and is the shortest-orbital-period system among all known UCBs with detected hydrogen but no detected helium in the spectrum. The evolved CV models typically predict that at 56\,min, most of the hydrogen should be stripped from the donor, and some helium should be visible. Only in models where the system is still early in its evolution and the donor is still warm and detectable (which is not the case here) would no helium be detectable. 
An alternative explanation is that this system has evolved from a detached white dwarf--brown dwarf binary \citep[which can reach periods as short as 37\,min before the start of accretion,][]{nelson2018,2021ApJ...913..118R}, and only started accretion close to the current orbital period, a scenario also suggested for the 66.6\,min orbital period cataclysmic variable SDSS~J1507 \citep{2007MNRAS.381..827L}.

\subsubsection{ZTF~J0654+3608}
The spectrum of ZTF~J0654+3608 does not resemble any other system in the sample. It only shows weak absorption lines, which is typical for AM~CVn systems that have a high accretion rate and ionised accretion disc (orbital periods of 10--22\,min). It has also never been seen to outburst over the 8.4\,yr ZTF baseline, which would also be consistent with a disc in a high accretion state. Note that it did trigger two ZTF alerts, but we attribute this to stochastic variability. 
Finally, the eclipse also appears broader (with a FWHM of 3\,min compared to a more typical eclipse width of 70\,s) and more `V'-shaped compared to the others, which would also suggest that the disc is bright. These three observational properties all suggest that this system has a high accretion rate.

The lines in the spectrum of ZTF~J0654+3608 are shallow, but a careful inspection of the spectrum shows a clear detection of hydrogen: four Balmer lines (H$\epsilon$, H$\delta$, H$\gamma$, and H$\beta$) are identified in absorption, with a combined significance that is stronger than that of \ion{He}{i}. This shows that the surface of the donor still contains a substantial amount of hydrogen. As already discussed in the case of ZTF~J2116+2446, this is unusual. As noted above, the minimum orbital period for a detached white dwarf--brown dwarf binary is 37\,min, almost exactly the orbital period of ZTF~J0654+3608. It is possible that ZTF~J0654+3608 evolved from a white dwarf--brown dwarf binary and only recently started accretion, which would also explain the high accretion rate. However, models of evolved CVs \citep{2023A&A...678A..34B} show that they can reach periods as short as 39\,min and still be dominated by hydrogen. 
Follow-up observations (to measure the donor mass, composition, and orbital period changes) would be valuable for testing UCB formation models. In particular, it would be very useful to measure the period change using the eclipses, to determine whether the orbital period is decreasing or increasing.

\subsubsection{AM~CVn binaries}
The other ten systems have properties that are very typical for AM~CVn binaries. Their periods of 26 to 65\,min are well within the traditional range and they have outburst frequencies that are (mostly) consistent with the outburst frequencies measured by \citet{levitan2015}. Of these ten objects, six have spectra that are not unusual compared to known AM~CVn systems. ZTF~J0859$-$1110 and ZTF~J1838+0744 show the strong helium emission of a bright disc, while ZTF~J1302$-$1746 and ZTF~J0151+0909 show helium emission with a narrow absorption core. The two shortest-period systems, ZTF~J0729$-$0602 and ZTF~J0701+5023, show weaker lines, as is typical at short orbital periods.

We do note that ZTF~J0531$-$2019 shows a single outburst, which is somewhat unusual: AM~CVn systems with such a long period generally do not outburst, while He~CVs typically show more frequent outbursts over an 8-yr baseline. This object could therefore be either an AM~CVn binary or a He~CV: its single outburst is consistent with either scenario, and a spectrum is needed to determine its exact nature. The other three systems without a spectrum, \object{ZTF~J0614+6014}, \object{ZTF~J0555$-$2345}, and ZTF~J1146$-$1649, have eclipse light curves and outburst behaviour that are fully consistent with being AM~CVn binaries, and we therefore classify them as such. Finally, we note that ZTF~J0859$-$1110 is the nearest known eclipsing AM~CVn binary, at a distance of 199\,pc.

\subsection{The sample of eclipsing white dwarf UCBs}
\begin{figure*}[t]
    \centering
    \includegraphics[width=\textwidth]{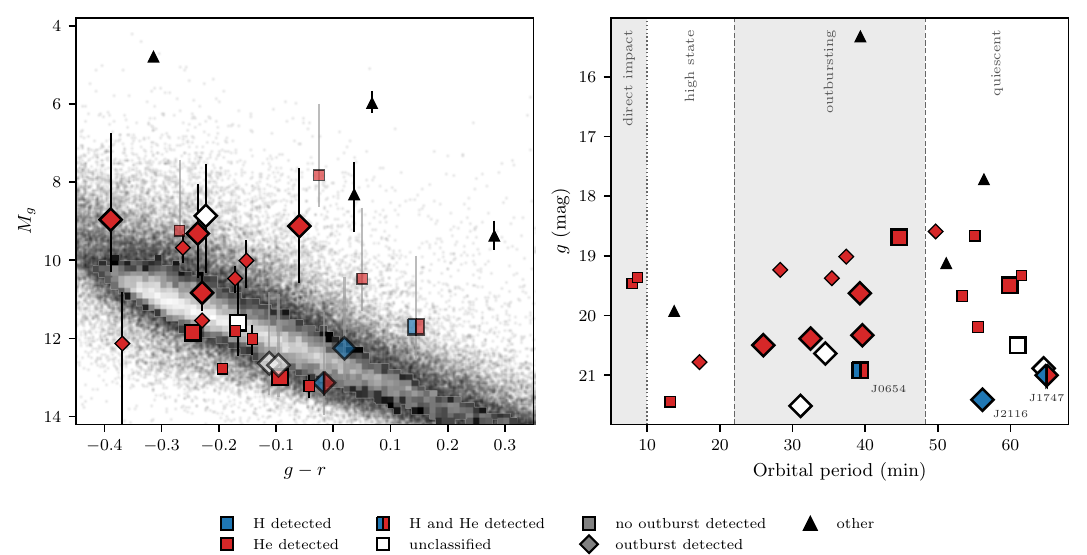}
    \caption{HR diagram (left) and orbital period vs.\ apparent $g$ magnitude (right) for the eclipsing accreting UCB sample. The grey background shows the general white dwarf population \citep{gentilefusillo2021}. Large markers show the objects from this paper, small markers show previously known systems. Marker colour indicates the spectroscopically detected species (hydrogen, helium, both, or white for no spectrum); marker shape indicates outburst status (diamond: outburst detected, square: none detected). Triangles mark objects where the donors are not (semi-)degenerate: the eclipsing WD+subdwarf-B binaries and the two warm-donor systems (ZTF~J1813+4251 and ZTF~J0127+5258). For objects without a \textit{Gaia} parallax measurement, we used a Galactic prior to estimate the distance, shown with transparent symbols in the left panel. In the right panel, the shaded bands and vertical lines mark the typical accretion
    states for AM~CVn binaries. Note that these are approximate, and not hard boundaries.}
    \label{fig:hr_period}
\end{figure*}

With this paper, we increase the number of known eclipsing accreting UCBs from 21 to 30. To put these newly discovered objects into context, Fig.~\ref{fig:hr_period} shows all the eclipsing accreting white dwarf UCBs. Systems without a \textit{Gaia} parallax are instead placed on this diagram using a Galactic exponential-density distance prior \citep{bailerjones2021}, with the scale height fit to this sample's own parallax-based distances rather than a fixed Galactic value. The HR diagram (Fig.~\ref{fig:hr_period}, left) shows that the new systems do not stand out relative to the previously known systems. We note that the three systems with detected hydrogen are redder than almost all the other eclipsing UCBs, ranking among the four reddest systems (in $g-r$) of the full sample. In addition, we also observe that the position of a system relative to the white dwarf sequence tracks its orbital period. Of the systems with a parallax-based distance, all those with $P_{\rm orb} \lesssim 40$\,min lie above the sequence (more luminous), by 1.3 to 3.1\,mag in $M_g$, while all those with $P_{\rm orb} \gtrsim 45$\,min lie on or slightly below it. The elevated systems are also the frequently outbursting ones. This is the expected behaviour: at shorter orbital periods the mass-transfer rate is higher, and the accretion disc rather than the accretor dominates the optical light, lifting the system above the cooling sequence. The long-period, non-outbursting AM~CVn systems with well-measured distances, in which the accretor dominates, instead sit slightly below the sequence. This is fully consistent with those accretors being smaller and more massive than typical field white dwarfs, which is in line with the measured masses of other (eclipsing) AM~CVn systems \citep[e.g.][]{vanroestel2022}.

The period-magnitude diagram (Fig.~\ref{fig:hr_period}, right) shows that the new objects are all relatively faint compared to the previously known sample (ZTF~J0151+0909 is the exception). In addition, our search method is particularly effective at finding outbursting, long-period systems. The newly discovered long-period objects are on average about 1.5\,mag fainter than the previously known, non-outbursting ones. If both populations have a similar intrinsic luminosity, this corresponds to roughly an order of magnitude more volume being probed, which suggests that the faint, long-period population is intrinsically about an order of magnitude rarer than the previously known, brighter but non-outbursting systems. Many more long-period, non-outbursting, deeply eclipsing systems therefore remain to be found if forced-photometry data are searched for all stars in ZTF.

\subsection{Gravitational wave emission and orbital period changes}
The evolution of UCBs is governed by gravitational waves. However, with \texttt{legwork} \citep{2022ApJS..260...52W} and assuming typical masses ($0.8$ and $0.05\,M_\odot$ for the accretor and donor), none of the 13 systems is individually detectable by LISA (S/N $>7$), even over a 10-yr mission. Nevertheless, within 10\,yr (before LISA launches) we can expect to measure a change in the eclipse arrival time of more than a second for systems with an orbital period less than 35\,min. We do note that the period change depends on the mass-transfer physics and the equation of state of the donor, and as \citet{2026MNRAS.548ag673G} showed, the long-period accreting UCBs do not always show a measurable period change. An eclipse timing campaign of these new systems can provide valuable insights into the physics of accreting compact binaries.

\section{Conclusions}
Nine accreting UCBs were discovered from their eclipses in ZTF photometry, and are presented here together with the four eclipsing systems of \citet{khalil2024}, analysed in the same way. Of the 13, ten are consistent with being AM~CVn binaries. 
Two systems, ZTF~J0654+3608 and ZTF~J1747+6302, show both hydrogen and helium in their spectra, an observational signature of the He~CV class, and likely evolved through the evolved CV channel. ZTF~J0654+3608 has a high accretion rate and likely hosts a high-state disc, despite its orbital period of 39.3\,min. The third hydrogen-rich system, ZTF~J2116+2446, shows only hydrogen lines and is the shortest-orbital-period system among all known UCBs with detected hydrogen but no detected helium. It may be an evolved CV, or it may descend from a white dwarf--brown dwarf binary that only recently started mass transfer. ZTF~J2116+2446 and ZTF~J1747+6302 also outburst every 2.8 and 0.3\,yr, respectively, far more frequently than AM~CVn binaries at their orbital periods of 56--65\,min, which supports a He~CV/CV nature.

The nine new systems increase the published sample of eclipsing accreting UCBs from 21 to 30\footnote{This count does not include the systems from \citet{2026arXiv261040346H}, published simultaneously.}. The new systems are mostly faint: the newly discovered long-period systems are on average about 1.5\,mag fainter than the previously known non-outbursting ones, which suggests that many more long-period, deeply eclipsing systems remain to be found in a search of ZTF forced photometry for all stars. None of the 13 systems is individually detectable by LISA, but for those with orbital periods below 35\,min, shifts in eclipse arrival times of more than a second should be measurable within a decade, providing a direct probe of their mass-transfer physics. Increasing the known sample of UCBs will be useful both for population studies and for detailed characterisation of individual objects, and both approaches are needed to better understand the UCB formation channels. High-speed photometry has already been obtained for most of the sample, and detailed characterisation will be presented in future work. The detailed characterisation of the three hydrogen-rich UCBs will be especially valuable in this respect.

\section*{Data availability}
The ZTF forced-photometry light curves and the reduced Keck/LRIS spectra of the 13 systems are available on Zenodo at \url{https://doi.org/10.5281/zenodo.22707181}.
The running census of all known eclipsing accreting white dwarf UCBs, from which the sample sizes quoted in this paper are taken, is maintained as a separate record at \url{https://doi.org/10.5281/zenodo.22706718}.

\begin{acknowledgements}
We thank the ZTF-collaboration referee, Igor Andreoni, for carefully reading the manuscript. We thank Boris G{\"a}nsicke and Keith Inight for comments on the paper.

This publication is part of the project ``The life and death of white dwarf binary stars'' (with project number VI.Veni.212.201) of the research programme NWO Talent Programme Veni Science domain 2021 which is financed by the Dutch Research Council (NWO).

Based on observations obtained with the Samuel Oschin Telescope 48-inch and the 60-inch Telescope at the Palomar Observatory as part of the Zwicky Transient Facility project. ZTF is supported by the National Science Foundation under Grants No. AST-1440341 (Phase 1) and AST-2034437 (Phase 2) and a collaboration including Caltech, IPAC, the Weizmann Institute for Science, the Oskar Klein Center at Stockholm University, the University of Maryland, Deutsches Elektronen-Synchrotron and Humboldt University, the TANGO Consortium of Taiwan, the University of Wisconsin at Milwaukee, Trinity College Dublin, Lawrence Livermore National Laboratories, and IN2P3, France. Operations are conducted by COO, IPAC, and UW.

The ZTF forced-photometry service was funded under the Heising-Simons Foundation grant \#12540303 (PI: Graham).

Some of the data presented herein were obtained at the W. M. Keck Observatory, which is operated as a scientific partnership among the California Institute of Technology, the University of California and the National Aeronautics and Space Administration. The Observatory was made possible by the generous financial support of the W. M. Keck Foundation. The authors wish to recognise and acknowledge the very significant cultural role and reverence that the summit of Maunakea has always had within the indigenous Hawaiian community. We are most fortunate to have the opportunity to conduct observations from this mountain. This work has made use of data from the European Space Agency (ESA) mission \textit{Gaia} (\url{https://www.cosmos.esa.int/gaia}), processed by the \textit{Gaia} Data Processing and Analysis Consortium (DPAC, \url{https://www.cosmos.esa.int/web/gaia/dpac/consortium}). Funding for the DPAC has been provided by national institutions, in particular the institutions participating in the \textit{Gaia} Multilateral Agreement.

This research made use of \texttt{numpy} \citep{numpy}, \texttt{scipy}
\citep{scipy}, \texttt{matplotlib} \citep{matplotlib}, \texttt{astropy}
\citep{astropy}, \texttt{emcee} \citep{emcee}, \texttt{pandas} \citep{mckinney2010},
\texttt{cuvarbase} \citep{hoffman2022}, and \texttt{legwork}
\citep{2022ApJS..260...52W}. The spectra were reduced with \texttt{lpipe}
\citep{perley2019}.

\textit{Use of AI tools.}
An AI coding assistant was used to draft and revise portions of the text. The authors take full responsibility for the content of this paper.
\end{acknowledgements}

\bibliographystyle{aa}
\bibliography{refs}

\begin{appendix}

\section{Ephemerides}
\label{app:ephemerides}

Table~\ref{tab:ephemerides} lists the orbital period and the epoch of
mid-eclipse of each system, as used to fold the light curves in
Fig.~\ref{fig:discovery_lcs}.

\begin{table}[h]
\centering
\caption{Ephemerides of the 13 eclipsing accreting UCBs. The orbital periods and
mid-eclipse epochs were determined by fitting a trapezoid-and-sine model to the folded ZTF
photometry. Times are barycentric Julian dates on the barycentric dynamical time scale.
Systems marked $^{a}$ were first reported by \citet{khalil2024}.}
\label{tab:ephemerides}
\small
\setlength{\tabcolsep}{2pt}
\begin{tabular}{lcccc}
\hline\hline
Name & $P_{\rm orb}$ & $\sigma(P_{\rm orb})$ & $T_0$ & $\sigma(T_0)$ \\
     & (d) & ($10^{-9}$ d) & (BJD$_{\rm TDB}$) & ($10^{-5}$ d) \\
\hline
ZTF~J0729$-$0602\tablefootmark{a} & 0.018045443 & 0.6 & 2459128.96408 & 2.9 \\
ZTF~J0614+6014 & 0.021600301 & 0.6 & 2459301.69278 & 2.9 \\
ZTF~J0701+5023\tablefootmark{a} & 0.022562838 & 0.3 & 2459179.88613 & 1.2 \\
ZTF~J0555$-$2345 & 0.023973617 & 0.9 & 2458472.86848 & 3.6 \\
ZTF~J1302$-$1746 & 0.027274689 & 0.4 & 2460112.70867 & 1.2 \\
ZTF~J0654+3608 & 0.027311237 & 2 & 2458869.79631 & 10.9 \\
ZTF~J1838+0744\tablefootmark{a} & 0.027508050 & 1 & 2458981.83585 & 4.2 \\
ZTF~J0151+0909 & 0.031039486 & 0.7 & 2459852.87769 & 1.4 \\
ZTF~J2116+2446\tablefootmark{a} & 0.038988890 & 1 & 2458802.62215 & 3.7 \\
ZTF~J0859$-$1110 & 0.041603940 & 1 & 2459226.39905 & 2.8 \\
ZTF~J1146$-$1649 & 0.042349932 & 4 & 2460020.87894 & 7.4 \\
ZTF~J0531$-$2019 & 0.044836759 & 2 & 2459586.73418 & 4.6 \\
ZTF~J1747+6302 & 0.045122312 & 1 & 2458893.08746 & 3.4 \\
\hline
\end{tabular}
\end{table}

\section{Line identifications}
\label{app:lines}

Table~\ref{tab:lines} lists every transition identified in the nine LRIS
spectra, system by system. These are the identifications made by eye on the
continuum-divided spectra (Sect.~\ref{sec:spectra}), and are the same ones
marked in Fig.~\ref{fig:spectra}.

\begin{table*}
\centering
\caption{Lines identified in the nine LRIS spectra. Rest wavelengths are in
\AA{} and a bullet marks each identification. The identifications are made by
eye on the continuum-divided spectra (Sect.~\ref{sec:spectra}) and are the same
ones marked in Fig.~\ref{fig:spectra}. Columns are ordered by increasing
orbital period and headed by the object name shortened to its first four
digits.}
\label{tab:lines}
\footnotesize
\setlength{\tabcolsep}{4pt}
\begin{tabular}{l r ccccccccc}
\hline\hline
Species & $\lambda$ & J0729 & J0701 & J1302 & J0654 & J1838 & J0151 & J2116 & J0859 & J1747 \\
\hline
He\,I & 3613.6 &  &  & $\bullet$ &  &  &  &  &  &  \\
He\,I & 3634.2 &  &  & $\bullet$ &  &  &  &  &  &  \\
He\,I & 3819.6 & $\bullet$ &  & $\bullet$ &  & $\bullet$ &  &  &  & $\bullet$ \\
He\,I & 3888.6 & $\bullet$ & $\bullet$ & $\bullet$ & $\bullet$ & $\bullet$ & $\bullet$ &  & $\bullet$ & $\bullet$ \\
He\,I & 4009.3 &  & $\bullet$ &  &  &  &  &  &  &  \\
He\,I & 4026.2 & $\bullet$ & $\bullet$ & $\bullet$ &  & $\bullet$ & $\bullet$ &  & $\bullet$ &  \\
He\,I & 4120.8 &  & $\bullet$ & $\bullet$ &  &  &  &  &  &  \\
He\,I & 4143.8 &  & $\bullet$ & $\bullet$ &  & $\bullet$ &  &  &  &  \\
He\,I & 4387.9 &  & $\bullet$ & $\bullet$ & $\bullet$ & $\bullet$ & $\bullet$ &  & $\bullet$ &  \\
He\,I & 4437.6 &  &  & $\bullet$ &  &  &  &  &  &  \\
He\,I & 4471.5 & $\bullet$ & $\bullet$ & $\bullet$ & $\bullet$ & $\bullet$ & $\bullet$ &  & $\bullet$ & $\bullet$ \\
He\,I & 4713.1 &  & $\bullet$ & $\bullet$ & $\bullet$ & $\bullet$ & $\bullet$ &  & $\bullet$ &  \\
He\,I & 4921.9 & $\bullet$ & $\bullet$ & $\bullet$ &  & $\bullet$ & $\bullet$ &  & $\bullet$ &  \\
He\,I & 5015.7 & $\bullet$ & $\bullet$ & $\bullet$ &  & $\bullet$ & $\bullet$ &  & $\bullet$ & $\bullet$ \\
He\,I & 5047.7 &  & $\bullet$ & $\bullet$ &  & $\bullet$ &  &  &  & $\bullet$ \\
He\,I & 5875.6 & $\bullet$ &  & $\bullet$ &  & $\bullet$ & $\bullet$ &  & $\bullet$ &  \\
He\,I & 6678.2 & $\bullet$ & $\bullet$ & $\bullet$ &  & $\bullet$ & $\bullet$ &  & $\bullet$ & $\bullet$ \\
He\,I & 7065.2 &  & $\bullet$ & $\bullet$ &  & $\bullet$ & $\bullet$ &  & $\bullet$ & $\bullet$ \\
He\,I & 7281.4 &  &  & $\bullet$ &  & $\bullet$ & $\bullet$ &  & $\bullet$ & $\bullet$ \\
He\,II & 4685.7 &  & $\bullet$ & $\bullet$ &  & $\bullet$ & $\bullet$ &  & $\bullet$ & $\bullet$ \\
H & 3970.1 &  &  &  & $\bullet$ &  &  &  &  &  \\
H & 4101.7 &  &  &  & $\bullet$ &  &  & $\bullet$ &  & $\bullet$ \\
H & 4340.5 &  &  &  & $\bullet$ &  &  & $\bullet$ &  & $\bullet$ \\
H & 4861.3 &  &  &  & $\bullet$ &  &  & $\bullet$ &  & $\bullet$ \\
H & 6562.8 &  &  &  &  &  &  & $\bullet$ &  & $\bullet$ \\
N\,I & 4099.0 &  &  &  &  &  & $\bullet$ &  &  &  \\
N\,I & 4109.0 &  &  &  &  &  & $\bullet$ &  &  &  \\
N\,I & 4137.0 &  &  &  &  &  & $\bullet$ &  &  &  \\
N\,I & 4151.0 &  &  &  &  &  & $\bullet$ &  &  &  \\
N\,I & 4214.5 &  &  &  &  &  & $\bullet$ &  &  &  \\
N\,I & 4223.5 &  &  &  &  &  & $\bullet$ &  &  &  \\
N\,I & 7423.0 &  &  &  &  &  & $\bullet$ &  &  &  \\
N\,I & 7442.0 &  &  &  &  &  & $\bullet$ &  &  &  \\
N\,I & 7468.0 &  &  &  &  &  & $\bullet$ &  &  &  \\
N\,II & 6482.0 &  &  &  &  &  & $\bullet$ &  &  &  \\
Ca\,II & 3933.7 &  & $\bullet$ & $\bullet$ & $\bullet$ & $\bullet$ & $\bullet$ & $\bullet$ &  & $\bullet$ \\
Ca\,II & 3968.5 & $\bullet$ & $\bullet$ & $\bullet$ &  & $\bullet$ & $\bullet$ & $\bullet$ &  & $\bullet$ \\
Na\,I & 5682.6 &  &  &  &  & $\bullet$ &  &  &  &  \\
Na\,I & 5688.2 &  &  &  &  & $\bullet$ &  &  &  &  \\
Na\,I & 5889.9 &  &  &  & $\bullet$ &  & $\bullet$ &  & $\bullet$ &  \\
Na\,I & 5895.9 &  &  &  & $\bullet$ &  & $\bullet$ &  & $\bullet$ &  \\
Si\,I & 3905.5 &  &  &  &  &  & $\bullet$ &  & $\bullet$ &  \\
Mg\,I & 3832.3 &  & $\bullet$ &  &  &  & $\bullet$ &  &  &  \\
Mg\,I & 3838.3 &  &  &  &  &  &  & $\bullet$ &  & $\bullet$ \\
Mg\,I & 5172.7 &  &  &  & $\bullet$ & $\bullet$ &  &  &  &  \\
Mg\,I & 5183.6 &  &  &  & $\bullet$ & $\bullet$ &  &  &  &  \\
Si\,II & 3856.0 &  &  &  &  &  & $\bullet$ &  & $\bullet$ &  \\
Si\,II & 3862.6 &  &  &  &  &  & $\bullet$ &  & $\bullet$ &  \\
Si\,II & 4128.1 &  &  & $\bullet$ &  &  & $\bullet$ &  &  &  \\
Si\,II & 4130.9 &  &  & $\bullet$ &  &  & $\bullet$ &  &  &  \\
Si\,II & 6347.1 &  &  & $\bullet$ &  & $\bullet$ & $\bullet$ &  &  & $\bullet$ \\
Si\,II & 6371.4 &  &  & $\bullet$ &  & $\bullet$ & $\bullet$ &  &  & $\bullet$ \\
\hline
\end{tabular}
\end{table*}

\section{Outburst light curves}
\label{app:outburst_lcs}

\begin{figure*}[hbtp]
    \centering
    \includegraphics[width=\textwidth]{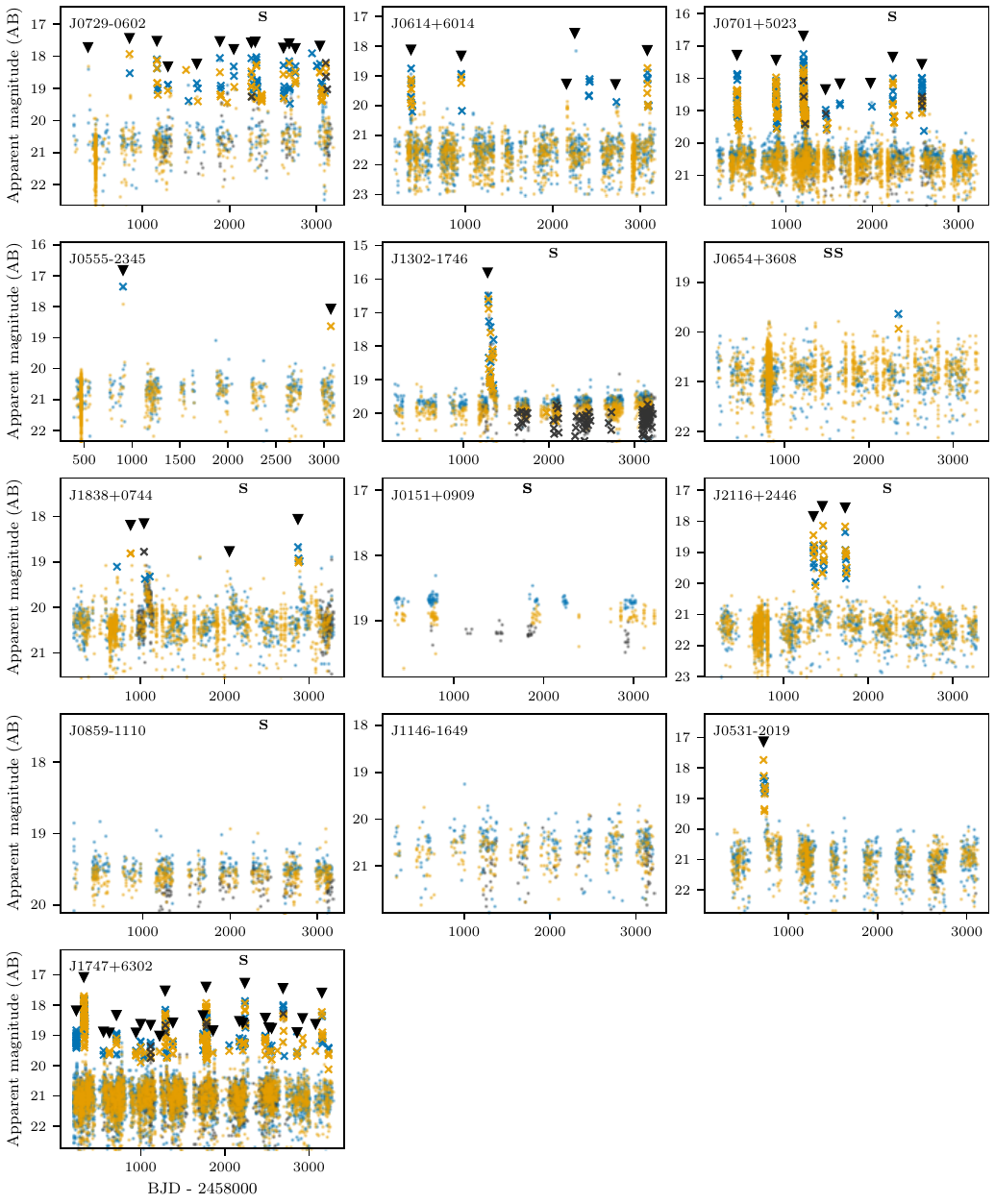}
    \caption{Unfolded survey light curves for the 13 eclipsing accreting UCBs. The data are shown as apparent magnitude against time (BJD $-$ 2458000). The crosses ($\times$) indicate ZTF alert photometry, while the dots represent the forced photometry. Blue, orange, and dark grey show ZTF $g$, $r$, and $i$, respectively. Black triangles mark the start of each confirmed outburst (Sect.~\ref{sec:outbursts}). A bold \textbf{S} marks each epoch at which an LRIS spectrum was taken (Sect.~\ref{sec:spectra}).}
    \label{fig:outburst_lcs}
\end{figure*}

Figure~\ref{fig:outburst_lcs} shows the full unfolded survey light curves for all 13 systems, highlighting the outburst behaviour (or lack thereof) across the entire observing baseline.

\end{appendix}
\end{document}